\documentclass[amsmath,amssymb,amsfonts,aps,pre,superscriptaddress,bibnotes,showpacs,showkeys,longbibliography,10pt,twocolumn,floatfix]{revtex4-2}
\usepackage{graphicx}
\usepackage{amsmath,amssymb,amsfonts}
\usepackage{sidecap}
\usepackage[english]{babel}
\usepackage{xcolor}
\usepackage{lipsum}
\usepackage{physics}
\usepackage{hyperref}

\newcommand{\me}{\mathrm{e}}

\usepackage{scalerel}
\usepackage{tikz}
\usetikzlibrary{calc}
\usetikzlibrary{patterns}
\usetikzlibrary{svg.path}
\definecolor{orcidlogocol}{HTML}{A6CE39}
\tikzset{
  orcidlogo/.pic={
    \fill[orcidlogocol] 
svg{M256,128c0,70.7-57.3,128-128,128C57.3,256,0,198.7,0,128C0,57.3,57.3,0,128,
0C198.7,0,256,57.3,256,128z};
    \fill[white] svg{M86.3,186.2H70.9V79.1h15.4v48.4V186.2z}
                 
svg{M108.9,79.1h41.6c39.6,0,57,28.3,57,53.6c0,27.5-21.5,53.6-56.8,
53.6h-41.8V79.1z 
M124.3,172.4h24.5c34.9,0,42.9-26.5,
42.9-39.7c0-21.5-13.7-39.7-43.7-39.7h-23.7V172.4z}
                 
svg{M88.7,56.8c0,5.5-4.5,10.1-10.1,10.1c-5.6,0-10.1-4.6-10.1-10.1c0-5.6,4.5-10.1
,10.1-10.1C84.2,46.7,88.7,51.3,88.7,56.8z};
  }
}

\newcommand\orcid[1]{\!%
  \href{https://orcid.org/#1}{%
    \mbox{%
      \scaleto{%
        \begin{tikzpicture}[yscale=-1,transform shape]
          \pic{orcidlogo};
        \end{tikzpicture}
      }{8pt}%
    }%
  }%
}

\makeatletter

\begin{document}

\graphicspath{ {Images/} }

\title{Trade-offs and thermodynamics of energy-relay proofreading}

\author{Jonas Berx~\orcid{0000-0003-1904-8103}}
\affiliation{Niels Bohr International Academy, Niels Bohr Institute, University of Copenhagen, Blegdamsvej 17, 2100 Copenhagen, Denmark}
\author{Karel Proesmans~\orcid{0000-0001-9051-4205}}
\affiliation{Niels Bohr International Academy, Niels Bohr Institute, University of Copenhagen, Blegdamsvej 17, 2100 Copenhagen, Denmark}
\date{\today}

\begin{abstract}
Biological processes that are able to discriminate between different molecules consume energy and dissipate heat. They operate at different levels of fidelity and speed, and as a consequence there exist fundamental trade-offs between these quantities and the entropy production rate. Usually, the energy source required to operate in a high-fidelity regime comes from the consumption of external energetic molecules, e.g., GTP hydrolysis in protein translation . In this work, we study trade-offs between several kinetic and thermodynamic observables for Hopfield's energy-relay mechanism, which does not consume external molecules and is able to operate in depleted regions, at the cost of a higher error rate. The trade-offs are obtained both analytically and numerically via Pareto optimal fronts. We find that the scheme is able to operate in three distinct regimes: an energy relay regime, a mixed relay-Michaelis-Menten regime, and a Michaelis-Menten regime, depending on the kinetic and energetic parameters that tune transitions between states. The mixed regime features a dynamical phase transition in the error-entropy production Pareto trade-off, while the pure energy relay regime contains a region where this type of proofreading energetically outperforms standard kinetic proofreading.
\end{abstract}

\maketitle

\section{Introduction \label{sec:intro}}
Life is a remarkable manifestation of non-equilibrium processes that rely on the precise handling of information. The intricate mechanisms involved in information processing within organisms are of seminal importance in the study of the physics of life. Among these mechanisms, the concept of kinetic proofreading (KPR), as introduced in the pioneering works of Hopfield and Ninio~\cite{Hopfield1974,Ninio1975}, emerged as a fundamental principle for achieving unexpected fidelity in, e.g., nucleic acid transcription and protein synthesis~\cite{Hopfield1976}. By cyclically resetting the chemical reaction system instead of proceeding to product formation, mistakes arising from the incorporation of wrong substrates can be rectified, leading to high accuracy. KPR achieves this goal by introducing a strongly irreversible step in these processes that relies on the hydrolysis of external molecules such as GTP.

One underlying assumption in the KPR scheme is that different production cycles are independent; the enzyme is reset to its unbound starting conformation after every product formation. When there are correlations present in the assembly process, as is the case in, e.g., enzymes that possess a single binding site with multiple internal states, proofreading properties can depend on the state of the enzyme after a previous production step was completed. Such time-dependent functionality is termed \emph{dynamic cooperativity}, contrasting with allosteric cooperativity~\cite{Qian2008}. Systems exhibiting dynamic cooperativity possess a mnemonic functionality; they can remember their history~\cite{Ninio1986,Job1988,Qian2008}. Such memory effects are relatively well-known in enzymology~\cite{CornishBowden1987}, but rarely in conjunction with the thermodynamics of accuracy~\cite{NINIO2006}.

In this work, we study a particular proofreading scheme that utilises such dynamic cooperativity: Hopfield's energy-relay proofreading (ERPR)~\cite{Hopfield1980}. ERPR offers an alternative approach to classical proofreading systems by eliminating the need to hydrolyse external molecules. For instance, in DNA replication, the initial formation of a new phosphodiester bond in the processive synthesis process liberates energy, transforming the enzyme into a high-energy, metastable state. This high-energy state is crucial for ensuring that the error-rejecting mechanism is irreversible. If errors had the potential to re-enter through the reverse of the pathway that initially rejected them, effective proofreading would be impossible. The correlations between consecutive assembly phases are thus inherently integrated into the internal chemical network. It has been established that free enzyme conformational fluctuations yield non-Michaelis-Menten (MM) kinetics and can lead to dynamic cooperativity \cite{Kumar2016}.

Currently, there is evidence pointing to the possible existence of ERPR in concrete biochemical systems. For instance, DNA polymerases exhibit both static and dynamic cooperativity~\cite{Hopfield1980}, further implying that they can exist in multiple functional (and presumably, conformational) states. In Ref.~\cite{Radding2004}, the authors consider a variant of the energy relay scheme to model the cavity system of myoglobin which protects it from carbon monoxide poisoning. More recently, it was suggested that RNA compression after the first CCA cycle of the CCA-adding enzyme is an example of proofreading via an energy relay mechanism~\cite{Kuhn2015}, where the potential energy stored in the enzyme due to compression provides the energy relay mechanism for proofreading in this case. This mechanism enables the enzyme to label structurally deficient tRNAs with a CCACCA-end, tagging them for subsequent degradation by exonuclease~\cite{Kuhn2015,Wilusz2015}.

Lastly, for DNA replication the initial insertion of nucleotides within each processive replication segment occurs with a notable propensity for errors in the energy relay mechanism \cite{Hopfield1980}, as a result of elevating the enzyme from a low-energy to a high-energy state, induced by the incorporation of a nucleotide through regular MM kinetics. This distinctive feature serves as a hallmark of the ERPR mechanism, providing a pathway for experimental validation. 

In this paper, we study the ERPR scheme in the setting of stochastic thermodynamics, to investigate the trade-off between speed, accuracy, proofreading cost and entropy production. Since the need for enhanced fidelity necessarily implies that living processes take place far from equilibrium, the trade-off between the error rate and entropy production rate will play a crucial role in our analysis. The KPR mechanism is one of the most extensively studied models in stochastic thermodynamics~\cite{Ehrenberg1980,Murugan2014,Sartori2015,Boeger2022}, since it offers a relatively simple testbed in which to compute analytically and validate experimentally different trade-offs~\cite{Bennett1979,Murugan2012,Rao_2015,Banerjee2017,Banerjee2020,Mallory2019,Yu2020} and bounds, such as the thermodynamic uncertainty relation~\cite{Barato2015,Pieros2020,berx2023}. In contrast, ERPR has not been studied within this setting, although both schemes offer a conducive way of proofreading. We consider in this work for simplicity a single-step scheme, although multi-step extensions have been studied as well~\cite{Savageau1981}.

The setup of this work is as follows. In Sec.~\ref{sec:ERPR_full} we set the stage for the energy relay mechanism and the associated observables, and briefly discuss the concept of Pareto optimal fronts. In Sec.~\ref{sec:energetic_kinetic}, we show that the ERPR scheme is able to operate in three distinct regimes, based on both the error rate and the cost of proofreading. We discuss these regimes in more detail in Sec.~\ref{sec:scaling_laws} and perform a scaling analysis of the pairwise Pareto optimal fronts. In Sec.~\ref{sec:ERPRvsKPR}, we directly compare the ERPR scheme with KPR in a model where the proofreading discrimination can also be varied by the Pareto optimisation, and finally we conclude in Sec.~\ref{sec:conclusions}.

\section{Energy relay proofreading and Pareto fronts\label{sec:ERPR_full}}
\begin{figure}[!htp]
    \centering
    \includegraphics{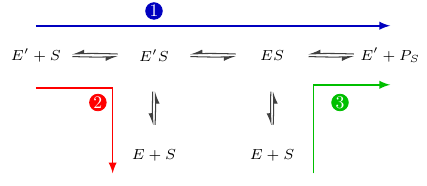}
    \caption{Reaction pathways for an enzyme $E$ and substrate $S\in\{R,W\}$ in the energy relay proofreading scheme. Branching pathways are indicated by different colours and numbers, where the arrows point in the preferential direction of the average flux.}
    \label{fig:ERPR_reaction}
\end{figure}

We start by setting the stage for the energy relay proofreading mechanism~\cite{Hopfield1980,Qian2010}. Consider an enzyme that can access energy states $E$ and $E'$, in order of increasing energy. We call henceforth $E$ the ground state and $E'$ the high-energy, metastable state. The enzyme can bind substrates $S \in\{R,W\}$, where we consider right ($R$) and wrong ($W$) substrates. The reaction pathways are given by the reaction scheme in Fig.~\ref{fig:ERPR_reaction}. Starting from its ground state, the enzyme $E$ can bind a substrate $S$ to form the complex $ES$, which can in turn produce the corresponding product $P_S$ by means of regular MM kinetics through pathway $\#3$. However, the energy released in the production process is used to elevate the enzyme to its high-energy state $E'$. This high-energy conformation can once again bind a substrate to form the complex $E'S$, which can either undergo proofreading by unbinding the substrate to form $E+S$ through pathway $\#2$, or can continue through the production pathway $\#1$.
In this scheme, the side branch $E'S \rightarrow E+S$ can be used to effectively proofread the complex. This is in stark contrast with KPR where only a single unbound enzymatic conformational state is considered and effective discrimination occurs by hydrolysing energetic molecules to irreversibly change the bound complex $ES$ to a state $ES^*$. From this state, the enzyme either unbinds the substrate, or proceeds to create product $P_S$. Both processes reset the enzyme to the unbound state, restarting the cycle. The probability of unbinding a wrong substrate is slightly higher than unbinding a right one, which is amplified through the resetting of the cycle, and therefore leads to effective proofreading

\begin{figure*}[htp]
    \centering
    \includegraphics[width=0.9\linewidth]{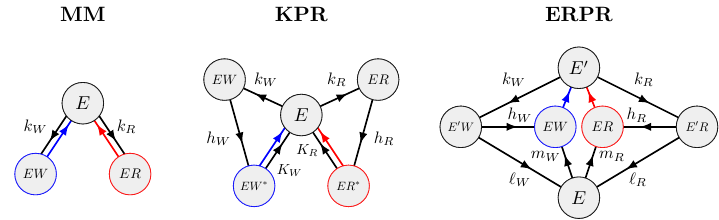}
    \caption{Discriminatory networks for the MM, KPR and ERPR schemes. Black edges in the networks are bidirectional, with the average direction of the flux indicated by arrows. Red/blue edges are unidirectional with catalysis rate $F$. }
    \label{fig:networks}
\end{figure*}

The full reaction network of both of these schemes, together with a standard MM network is shown in Fig.~\ref{fig:networks}. There are some important differences between ERPR and other proofreading schemes such as KPR. Firstly, ERPR generally works without consuming energy in, e.g., NTP hydrolysis. Secondly, the presence of multiple unbound enzymatic states in ERPR allows for correlations between subsequent cycles.

\subsection{Kinetic and thermodynamic observables}
To continue, we chemostat the substrate and product concentrations $[R],[W],[P_R]$ and $[P_W]$, such that the time evolution of the concentrations $[\xi]$ of the different enzymatic configurations $\xi\in\{E',E,E'R,E'W,ER,EW\}$ is given by the chemical master equation (CME):

\begin{equation}
    \label{eq:master_eq_concentrations}
    \dv[]{\Xi}{t} = {\bf \Gamma}\,\Xi\,,
\end{equation}
where $\Xi(t)$ denotes a vector of concentrations of the different states $\xi$ at time $t$, and the matrix ${\bf \Gamma}$ contains the transitions between states. In particular, for the ERPR the master equation is given by a set of six coupled differential equations, c.f., Eqs.~\eqref{eq:master_eq_ERPR} below.

Additionally, we require that the total concentration of enzymes is fixed and normalised to one, i.e., $[E']+[E] + [E'R] + [E'W] + [ER] + [EW] = 1$. Consequently, we can interpret all of our results as if a single enzyme cycles through the different states. The master equation then becomes 
\begin{equation}
    \label{eq:master_eq_general}
    \dv[]{\mathbb{P}}{t} = {\bf \Gamma}\,\mathbb{P}\,,
\end{equation}
where $\mathbb{P}(t)$ denotes now the vector of probabilities $P_\xi(t)$ to find the enzyme in state $\xi$ at time $t$. The steady state probabilities $P_\xi^s$ of the system are obtained by setting the left-hand side of equation~\eqref{eq:master_eq_general} to zero. The final catalysis step, whereby the products are formed and the enzyme is brought back to its high-energy state $E'$, is chosen to be unidirectional with rate $F$, which is equal for both right and wrong product formation. This last assumption is made for mathematical simplicity and because this process is highly irreversible in practice. Consequently, we can depict consecutive occurrences of the process by disregarding their final product state and treating this reaction as an irreversible pathway leading back to the high-energy free-enzyme state.

The kinetic quantities that we are interested in are the error rate $\eta$, proofreading cost $C$ and production speed $v$. We first consider the error rate $\eta$, defined as the ratio of the wrong production flux $\mathcal{J}_W$ and the total production flux $\mathcal{J}_p \equiv \mathcal{J}_W + \mathcal{J}_R$,
\begin{equation}
    \label{eq:error_rate}
    \eta = \frac{\mathcal{J}_W}{\mathcal{J}_R + \mathcal{J}_W}\,,
\end{equation}
where the steady-state production fluxes are defined as $\mathcal{J}_{S} \equiv F\,P_{ES}^s$, $S\in\{R,\,W\}$.

We next consider the cost of proofreading $C_j$ via either pathway $j=2$ or $j=3$ (see the red and green lines in Fig.~\ref{fig:ERPR_reaction}, respectively), where a bound enzyme, either $E'S$ or $ES$, releases its substrate $S$ and changes to the ground state $E$. These costs of proofreading are defined as
\begin{equation}
    \label{eq:proofreading_cost_2}
    \begin{split}
    C_2 &= \frac{\mathcal{J}_{\ell,R}+\mathcal{J}_{\ell,W}}{\mathcal{J}_p}\,, \\
    C_3 &= -\frac{\mathcal{J}_{m,R}+\mathcal{J}_{m,W}}{\mathcal{J}_p}\,,
    \end{split}
\end{equation}
with $\mathcal{J}_{\ell,S} = \ell_S^+ P_{E'S}^s - \ell_S^- P_E^s$ and $\mathcal{J}_{m,S} = m_S^+ P_{E}^s - m_S^- P_{ES}^s$ for $S\in\{R,\,W\}$. The reason for calling this quantity a `cost' derives from the equivalent nomenclature used in studies on KPR~\cite{Yu2022}, where the number of times the proofreading side reaction is used is related to the amount of NTP that is hydrolysed. Therefore, in studies on ERPR~\cite{Savageau1981} and hence also in this work, this quantity is referred to as a cost. Despite no longer directly being related to the dissipated energy for ERPR, $C_j$ can still provide important insights. Indeed, it can be seen as the number of times the proofreading pathways $\#2$ or $\#3$ are used per product molecule formed, and this is crucial to determine the different operating regimes of ERPR as we will see below.

In the steady state, the fluxes into state $E$ balance and the total proofreading cost becomes zero, $C = C_2 + C_3 = 0$. It can be shown that the individual pathway costs are bounded by $|C_i|\leq 1$, with $i=2,3$, when $\mathcal{J}_{h,R}+\mathcal{J}_{h,W} \geq 0$, with $\mathcal{J}_{h,S} \equiv h_S^+ P_{E'S}^s - h_S^- P_{ES}^s$. Since this step is usually assumed to be irreversible for high fidelity regimes~\cite{Hopfield1980,Savageau1981}, as is also done for KPR schemes~\cite{Hopfield1974}, we assume that this condition is fulfilled as well in the regimes of the trade-offs we will study. While the total cost may be zero, it is instructive to study the partial proofreading costs themselves, i.e., equation~\eqref{eq:proofreading_cost_2}. Since these quantities can in principle both take on either positive or negative values, they can be used to study the fraction of products that have passed through the proofreading cycle, either in the forward or backward direction. Following the definition of the cost in KPR, we consider here only the partial cost associated with pathway $\#2$, i.e., $C_2$, which we will henceforth just call the cost $C$.

The last kinetic quantity we are interested in is the production speed $v$, which is defined as the dimensionless ratio of the production flux and the catalysis rate, i.e.,
\begin{equation}
    \label{eq:speed}
    v = \mathcal{J}_p/F\,.
\end{equation}
where we divide by $F$ to make the speed dimensionless.

Finally, following~\cite{Rao_2015,Chiuchiu_2023} we consider the thermodynamics of proofreading through the entropy production 
\begin{equation}
    \label{eq:entropy_production}
    \sigma = \frac{1}{2}\sum_{i,j}\,'\left(k_{ij}P_i^s-k_{ji}P_j^s\right) \ln\frac{k_{ij}P_i^s}{k_{ji}P_j^s}\,,
\end{equation}
where the primed sum runs over all pairs of states, not taking the catalysis transitions into account, and the $k_{ij}$ indicate the transition rates from state $i$ to $j$. The term between brackets can be seen as the probability flux between two states, while the logarithm is the affinity. It is clear that the catalysis transitions merely describe the succession of subsequent production steps and do not constitute any discrimination. It can be thought of as a separately optimised process. Therefore we do not take these steps into account for the entropy production. The entropy production per catalysis step is then given by $\Delta\sigma = \tau\sigma = \sigma/\mathcal{J}_p$, which is proportional to the free energy irreversibly lost during product formation.

\subsection{Pareto optimal fronts}

By tuning the kinetic rates within a feasible solution space, observables that depend directly on these rates become indirectly coupled to each other. As a result, optimisation of a single objective cannot generally be performed independently of the others. For the ERPR, we consider the set of feasible combinations $\mathcal{F} = (v, \Delta\sigma,C,\eta)$ of the observables defined earlier: speed $v$, entropy production rate per product $\Delta\sigma$, proofreading cost $C$ and error rate $\eta$). The goal is to improve these observables (or objectives), where improving an objective in our system corresponds to increasing the speed and decreasing the entropy production rate, cost and error. By means of multi-objective optimisation, the Pareto front $\mathcal{P}\subset \mathcal{F}$ can be computed, which represents the most optimal trade-off between the different targets. In essence, the Pareto front $\mathcal{P}$ defines a set of solutions where improving one objective necessarily comes at the detriment of another. Pairwise trade-offs between two objectives can then be determined by marginalising the concomitant four-dimensional Pareto optimal front.

\section{Energetic vs. kinetic discrimination \label{sec:energetic_kinetic}}
\begin{figure*}[htp]
    \centering
    \includegraphics[width=0.85\linewidth]{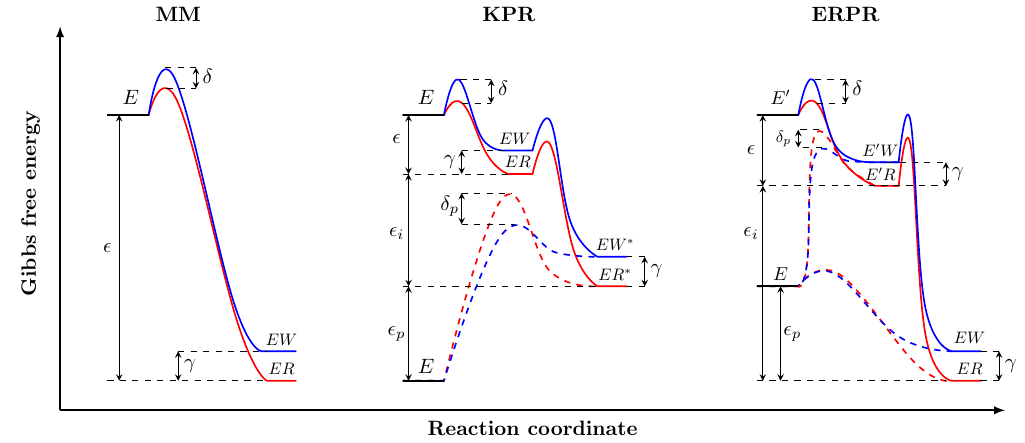}
    \caption{The Gibbs free energy ($G$) landscape as a function of the reaction coordinate for the MM, KPR and ERPR discriminatory mechanisms. Full red (blue) curves indicate the free energy landscape for the right (wrong) production pathways. Dashed lines indicate side reaction pathways, if any. For the ERPR scheme, two side reaction pathways (proofreading and upcycling) are drawn.}
    \label{fig:energy_landscape}
\end{figure*}

To continue, we will make a particular choice for the kinetic rates based on the free energy landscape shown in the right panel of Fig.~\ref{fig:energy_landscape}. For comparison, the free energy landscapes for the MM and KPR schemes are also shown~\cite{Rao_2015}. These rates offer the possibility of \emph{energetic} or \emph{kinetic} discrimination~\cite{Sartori2013}. The former indicates that discrimination results from an intrinsic difference in affinity of the right and wrong substrates, i.e., it is a result of the energy difference between the different enzymatic states. The latter mechanism pertains to kinetic barriers between different enzymatic states. This can be separated into two types of discrimination: i) \emph{initial} kinetic discrimination, where the kinetic barrier  occurs in the binding transitions from the high-energy free enzyme $E'+S$ to the $E'S$ states, or ii) \emph{proofreading} kinetic discrimination, which occurs between transitions from the bound $E'S$ states to the free enzymatic ground state $E$. For initial discrimination, the activation energy difference between the wrong and right pathways is given by $\delta \geq 0$, which we call the \emph{initial} discrimination parameter. Similarly, for proofreading discrimination, the difference between the right and wrong activation energy is given by $\delta_p \geq 0$, which we name the \emph{proofreading} parameter. Finally, the difference in binding energies between the two final bound states $ER$ and $EW$ is given by $\gamma$, the \emph{energetic} discrimination parameter.

The kinetic rates for the ERPR are given by the following Kramers rates,
\begin{alignat}{2}
k_R^+ &= \omega \me^{\epsilon+\delta},  &\qquad  k_R^- &= \omega \me^\delta,\nonumber \\  k_W^+ &= \omega \me^{\epsilon}, &\qquad k_W^- &= \omega \me^\gamma,\nonumber\\
h_R^+ &= \omega_i \me^{\epsilon_i},  &\qquad  h_R^- &= \omega_i,  \nonumber\\  h_W^+ &= \omega_i \me^{\epsilon_i}, &\qquad h_W^- &= \omega_i,\label{eq:rates}\\
\ell_R^+ &= \omega_\ell \me^{\epsilon_i-\epsilon_p-\delta_p},  &\qquad  \ell_R^- &= \omega_\ell \me^{-\delta_p}, \nonumber \\  \ell_W^+ &= \omega_\ell \me^{\epsilon_i-\epsilon_p+\gamma}, &\qquad \ell_W^- &= \omega_\ell, \nonumber\\
m_R^+ &= \omega_m \me^{\epsilon_p},  &\qquad  m_R^- &= \omega_m,  \nonumber\\  m_W^+ &= \omega_m \me^{\epsilon_p-\gamma}, &\qquad m_W^- &= \omega_m\nonumber\,.
\end{alignat}

It is straightforward to check that with this choice of kinetic rates, the Gibbs free energy difference between the right and wrong pathways is given by $\Delta G = \Delta \mu_R - \Delta \mu_W = \gamma$, with $\Delta\mu_R \equiv \mu_{P_R} - \mu_R$ and $\Delta\mu_W \equiv \mu_{P_W} - \mu_W$. Moreover, the chemical potential difference in the proofreading and upcycling cycles is equal to zero.

Motivated by experiments showing that tRNA discrimination of the CCA-adding enzyme is kinetically controlled \cite{DUPASQUIER2008,Hou2010}, we focus here on a \emph{forward} discrimination strategy, where the kinetic differences are present in the forward rates. This is a choice that pertains in particular to error correction mechanisms used by, e.g., ribosomes~\cite{Banerjee2017,Rodnina2017,Gromadski2004,Ehrenberg1980}, where discrimination of cognate and non-cognate aa-tRNA occurs in two phases: initial selection and proofreading. In both ribosomal selection steps, discrimination is based on different stabilities of correct and incorrect codon–anticodon duplexes, which manifests itself in a difference of forward kinetic rates~\cite{Rodnina2001}. 

Similarly, one can opt for discrimination in the reverse rates~\cite{Chiuchiu_2023}, which leads to qualitatively similar results. 

With this choice of rates, the discrimination error achieved simply by binding substrates and creating products is given by the ratio of the Boltzmann weights of the states of the bound enzyme, which simplifies to
\begin{equation}
    \label{eq:error_eq}
    \eta_{\rm eq} = \frac{1}{1+\me^\gamma}\,.
\end{equation}

We now introduce the shorthand notation $f(x) = 1/(1+\me^x)$, in order to ease notation where necessary. Hence, the error becomes $\eta_{\rm eq} = f(\gamma)$. Solving the master equation,
\begin{widetext}
\begin{equation}
    \label{eq:master_eq_ERPR}
    \begin{split}
        \dv[]{[E']}{t} &= k_R^- [E'R] + k_W^- [E'W] + F ([ER]+[EW])-(k_R^+[R]+k_W^+[W])[E']\\
        \dv[]{[E'S]}{t} &= k_S^+[S] [E'] + h_S^- [ES] + \ell_S^-[S][E] - (k_S^- + h_S^+ + \ell_S^+)[E'S]\,,\\
        \dv[]{[ES]}{t} &= h_S^+ [E'S] + m_S^+ [S][E] -(F+h_S^- +m_S^-) [ES]\,, \\
        \dv[]{[E]}{t} &= \ell_R^+ [E'R] + \ell_W^+ [E'W] + m_R^- [ER] + m_W^- [EW] - \{(\ell_R^- + m_R^+) [R] + (\ell_W^- +m_W^+) [W]\}\, [E]\,,
    \end{split}
\end{equation}
\end{widetext}
for $S\in\{R,\,W\}$ and subsequently normalising yields the steady-state distribution for the occupation probabilities, from which all of the observables can be calculated. While the system~\eqref{eq:master_eq_ERPR} is exactly solvable, the complete solution is long and unwieldy and hence we will not give the full expressions for the steady-state solution. However, analytical progress can be made when considering the limit in which the production cycles for both $R$ and $W$ are strongly driven in the forward direction, following pathway~$\#1$ in Fig.~\ref{fig:ERPR_reaction}. Additionally, we expect that effective proofreading is possible whenever the proofreading cycles are driven in the direction $E'S\rightarrow E+S\rightarrow ES$. 

Solving the master equation~\eqref{eq:master_eq_ERPR}, inserting the steady-state probabilities into~\eqref{eq:error_rate} and subsequently taking the limit of all the reverse kinetic rates to zero, yields the following expression for the error in the forward driven limit,
\begin{equation}
    \label{eq:error_forward_full}
    \eta = \frac{(\mathcal{M}-1) \Gamma_R +\Gamma_W \left(1+\mathcal{K} (\Gamma_R -1)\right)}{\mathcal{K} \mathcal{M} \Gamma_R \Gamma_W}\,,
\end{equation}
with $\Gamma_S = 1+\ell_S^+/h_S^+$, $\mathcal{K} = 1+k_R^+/k_W^+$ and $\mathcal{M} = 1+m_R^+/m_W^+$. Substituting the rates~\eqref{eq:rates}, the error in this forward driven limit reduces to 
\begin{equation}
    \label{eq:error_forward}
    \eta = f(\delta)f(\gamma)\left\{1+\frac{\me^{\delta-\delta_p}}{\frac{\omega_i}{\omega_\ell} \me^{\epsilon_p} + \me^{-\delta_p}} + \frac{\frac{\omega_i}{\omega_\ell}\me^{\epsilon_p +\gamma}}{\frac{\omega_i}{\omega_\ell} \me^{\epsilon_p} + \me^\gamma}\right\}\,.
\end{equation}

The free parameters that can be optimised are $\omega_i,\,\omega_\ell$ and $\epsilon_p$, so we introduce the (degenerate) variable $x\equiv \frac{\omega_i}{\omega_\ell}\me^{\epsilon_p}$. Essentially, $x$ is the ratio of the forward right transition rate through pathway~$\#1$ and the proofreading pathway~$\#2$ for a given value of $\delta_p$, i.e., $x=\frac{h_R^+}{\ell_R^+}\me^{-\delta_p}$. The minimal proofreading error $\eta_{\rm min,ERPR}$ can be found by minimising~\eqref{eq:error_forward} with respect to $x$. This yields
\begin{equation}
    \label{eq:error_minimal}
    \eta_{\rm min,ERPR} = f(\gamma)f(\delta)\frac{\me^{\gamma+\delta_p} + 2\me^{\gamma + \frac{\delta+\delta_p}{2}}-(1+\me^\gamma + \me^\delta) }{\me^{\gamma+\delta_p}-1}\,,
\end{equation}
which is achieved when $x = x^* \equiv (\me^{\delta/2} - \me^{-\delta_p/2})/(\me^{\delta_p/2} - \me^{\delta/2-\gamma})$. It can be proven (see appendix~\ref{app:appendix_B}) that the minimal error of the energy relay is always smaller than or equal to the minimal MM error~\cite{Rao_2015},
\begin{equation}
    \label{eq:error_minimal_MM}
    \eta_{\rm min,MM} = f(\max\{\delta,\gamma\})=\frac{1}{1+\me^{\max\{\delta,\gamma\}}}\,.
\end{equation}

To simplify notation, we will define $\eta_E \equiv \eta_{\rm min,\,ERPR}$ and $\eta_M \equiv \eta_{\rm min,\,MM}$ as the minimum error rates in both models. For the sake of completeness, we also give the minimal error in the KPR scheme, i.e.,
\begin{equation}
    \label{eq:error_minimal_KPR}
    \eta_{K} = f(\max\{\delta,\gamma\}+\gamma+\delta_p)=\frac{1}{1+\me^{\max\{\delta,\gamma\}+\gamma+\delta_p}}\,,
\end{equation}
although care should be taken when directly comparing the minimal errors of KPR and ERPR; the former possesses a proofreading step with the same kinetic discrimination parameter $\delta_p$ as the ERPR but the proofreading occurs from the bound states $ES^*$ and subsequently resets the enzyme to its unbound state, while for the ERPR scheme proofreading occurs from the high-energy bound states $E'S$ and then accesses the free enzyme ground state instead of resetting the system. We make a comparison between the ERPR and KPR in sec.~\ref{sec:ERPRvsKPR}.

By inspecting the expression for $x^*$, one can show that when $x^* \rightarrow\infty$, the minimal ERPR error becomes equal to the minimal MM error in the kinetic regime $\eta_M = f(\delta)$. For $x^*$, this means that when 
\begin{equation}
    \label{eq:condition_zero_cost}
    \delta = 2\gamma+\delta_p\,,
\end{equation}
the discriminatory mechanism corresponds to pure MM kinetics.

Indeed one can verify that $\gamma = (\delta - \delta_p)/2$ into equation~\eqref{eq:error_minimal}, leads to $\eta_{E} = f(\delta) = \eta_M$. The proofreading scheme stops using pathways~$\#2$ and~$\#3$ altogether for minimal $\eta$, but creates products only through pathway~$\#1$, where only the high-energy free enzymatic state $E'$ is used. This can be seen by inspecting $x$, which can only diverge when the ratio of kinetic rates $h_R^+/\ell_R^+$ diverges. At the transition characterised by~\eqref{eq:condition_zero_cost}, this divergence is associated with the limits $\ell_R^+\downarrow 0$ and $h_R^+\uparrow \infty$, effectively reducing the scheme to MM kinetics. 

To illustrate this transition between proofreading regimes in more detail, let us now turn to the proofreading cost $C$ in the forward limit, using the same rates as before, i.e.,
\begin{equation}
    \label{eq:cost_forward}
    C = f(\delta)\frac{1+\me^\delta + x(\me^{\delta-\gamma}+\me^{\delta_p})}{\left(1+x\me^{\delta_p}\right)\left(1+x\me^{-\gamma}\right)}\,.
\end{equation}

By eliminating $x$ in equations~\eqref{eq:error_forward} and~\eqref{eq:cost_forward}, it is possible to compute the relationship between the cost and the error in the forward driven system; this calculation is done in the appendix~\ref{app:appendix_C}, resulting in~\eqref{eq:cost_error_tradeoff}. It will turn out that this expression constitutes the optimal trade-off between those two quantities. 

For $\eta = \eta_{\rm eq}$, the proofreading cost becomes 
\begin{equation}
    \label{eq:cost_forward_eq}
    C(\eta_{\rm eq}) = \frac{f(\delta)}{f(\delta_p)}\,\frac{(\me^\gamma - \me^\delta)}{(\me^{\gamma+\delta_p}-1)}\,.
\end{equation}
Note that the proofreading cost at $\eta_{\rm eq}$ becomes zero at $\delta=\gamma$, i.e., on the transition to the kinetic discrimination regime \cite{Sartori2013}. When the error is minimal at $x=x^*$, the corresponding proofreading cost becomes
\begin{equation}
    \label{eq:cost_forward_minimal}
    C(\eta_{E}) = f(\delta)\frac{(\me^{\gamma+\delta_p/2}-\me^{\delta/2}) \left(\me^{\delta/2} + \me^{\delta_p/2}\right) }{\me^{\gamma+\delta_p}-1}\,.
\end{equation}

This minimal proofreading cost can also become zero or negative, the latter of which indicates that the proofreading cycle is being operated in reverse; substrate $S$ is bound to the enzyme $E$ in the ground state and forms a high-energy complex $E'S$. This particular transition will essentially never occur, but nevertheless we have taken the possibility into account by choosing $\ell^\pm_S$ to obey local detailed balance, contrary to the original definition of Hopfield~\cite{Hopfield1980} and others~\cite{Savageau1981}. The minimal cost is smaller than or equal to zero when the bound~\eqref{eq:condition_zero_cost} is fulfilled. 

\begin{figure}[htp]
    \centering
    \includegraphics[width=0.7\linewidth]{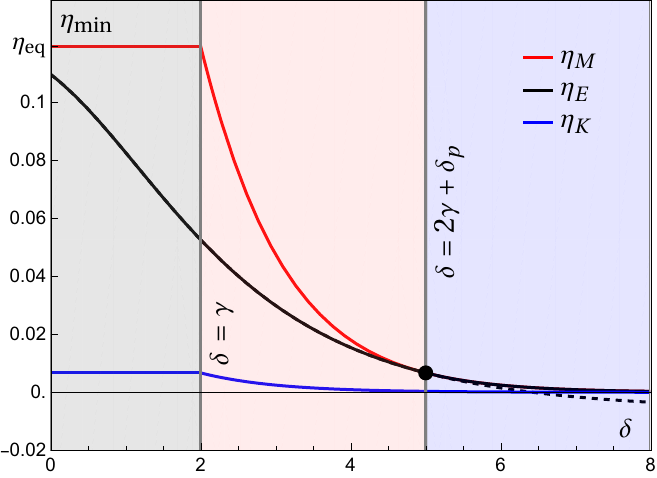}
    \caption{The minimal discrimination errors, $\eta_M$ (red), $\eta_E$ (black) and $\eta_K$ (blue), as a function of the discrimination parameter $\delta$, with $\gamma=2$ and $\delta_p=1$. The three discrimination regimes, ERPR, ERPR+MM and MM, are shown as grey, red and blue shaded areas, respectively. The black dashed line indicates the continuation of $\eta_E$ in the MM region.}
    \label{fig:errors_vs_delta}
\end{figure}

In Fig.~\ref{fig:errors_vs_delta}, we compare the minimal errors $\eta_E$, $\eta_M$ and $\eta_K$ as functions of $\delta$, with fixed $\gamma=2,\,\delta_p=1$. It can  be seen that increasing $\delta$ heralds a first transition at $\delta = \gamma$, to a proofreading mode where the system can initially make use of MM kinetics in the kinetic discrimination regime, for errors that are higher than the minimal MM error rate. For errors smaller than the minimal MM error rate, the proofreading side chains are used. Therefore, the minimal error rate in this mixed regime is set by the minimal ERPR error rate. A more detailed analysis of this mixed proofreading regime will be performed in Sec.~\ref{sec:scaling_laws}. The second transition at $\delta = 2\gamma + \delta_p$, set by~\eqref{eq:condition_zero_cost} effectively equates the minimal MM and ERPR errors, after which the MM error is preferred, as a result of $x$ becoming zero and the scheme transitioning fully into the MM regime. 

The three minimal errors obey a particular ordering, i.e., $\eta_K \leq \eta_E \leq \eta_M$, see Fig.~\ref{fig:errors_vs_delta}. We prove this in the appendix~\ref{app:appendix_B}.

\section{Pareto optimal trade-offs and scaling laws}\label{sec:scaling_laws}
We now study the trade-offs between the different quantities, i.e, $\eta,\,v,\,\Delta\sigma$ and $C$, as a function of the discrimination constants $(\gamma,\delta,\delta_p)$, by means of Pareto optimal fronts. For simplicity, we fix $\gamma=2$, $\epsilon=5$ (in units of $k_B T$) and $\omega = 1$. After choosing particular values for $\delta,\,\delta_p$, the number of remaining variables with respect to which needs to be optimised is equal to six.

To discover the Pareto optimal solutions, we employ a computational approach rooted in multifunction optimisation. Specifically, our methodology involves a thorough traversal of parameter space utilising genetic algorithms, using Matlab's \texttt{gamultiobj} function~\cite{Chiuchiu_2023}. We complement the optimisation algorithm with the constraint $\epsilon_p \leq \epsilon_i$, in order to maintain the ordering of the energy states as shown in Fig.~\ref{fig:energy_landscape}.

\begin{figure*}[htp]
    \centering
    \includegraphics[width=\linewidth]{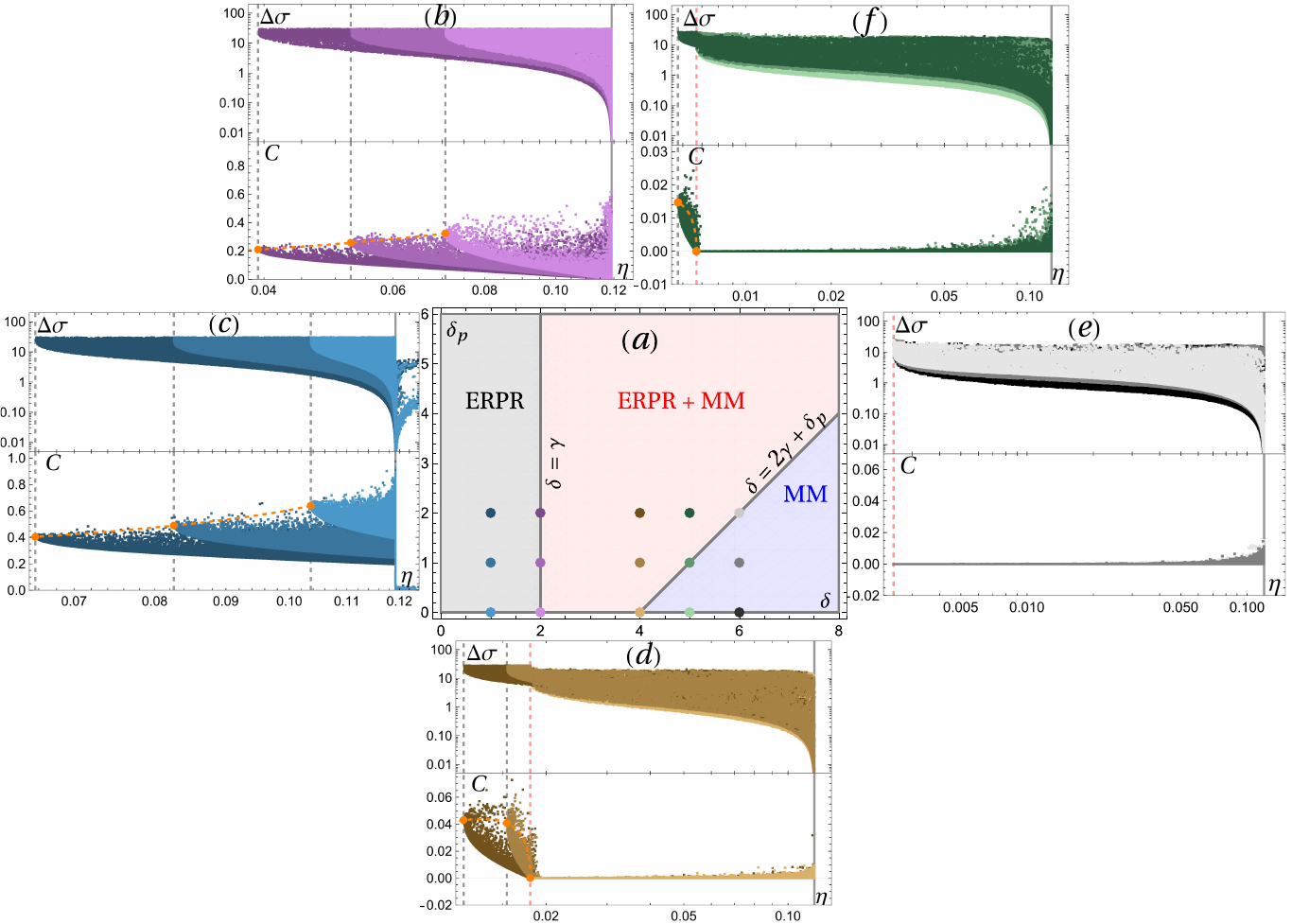}
    \caption{{\bf (a)} $(\delta,\delta_p)$ phase space for the ERPR scheme with $\gamma=2$. The three proofreading regimes are shown, together with the $(\delta,\,\delta_p)$ combinations (coloured points) that are studied in more detail in the other panels. {\bf (b-f)} Projected Pareto fronts onto the $(\Delta\sigma,\eta)$ and $(C,\eta)$ subspaces, for $\delta=1$ (blue), $\delta=\gamma=2$ (purple), $\delta=4$ (orange), $\delta=5$ (green) and $\delta=6$ (gray). The grey full, grey dashed, and red dashed vertical lines indicate respectively analytical predictions for $\eta_{\rm eq}$, $\eta_E$ and $\eta_M$. Orange dashed lines indicate a parametric plot of $C(\eta_E)$ vs. $\eta_E$, where $\delta$ is varied; corresponding orange dots are the $(\eta_E,C(\eta_E))$ values for the chosen parameters.}
    \label{fig:Paretogrid}
\end{figure*}

Based on the condition~\eqref{eq:condition_zero_cost}, we can draw the general proofreading behaviour in the $(\delta,\delta_p)$ plane, see Fig.~\ref{fig:Paretogrid}(a). For an initial discrimination factor $\delta\leq\gamma$ and $\delta_p\geq0$, i.e., the energetic discrimination regime, the system favours the energy relay to decrease the error fraction (grey shaded area). When $\delta\geq\gamma$, however, the system uses MM kinetics for error rates that are higher than the minimal MM error rate.  
When the minimal MM error rate is reached in this regime (red shaded area), the energy relay side chain is activated to achieve an even lower minimal error, given by equation~\eqref{eq:error_minimal}. This dual mechanism completely reduces to MM discrimination when $\delta\geq2\gamma+\delta_p$ (blue shaded area).

To find trade-offs between pairs of variables, we numerically marginalise the full four-dimensional objective space of the Pareto front, projecting onto the subspace under scrutiny. We will consider the different trade-offs in the three aforementioned regimes separately. We find that the production speed $v$ in all regimes remains equal to its value of $v=1$ at $\eta_{\rm eq}$, for error values far below $\eta_{\rm eq}$. This signifies that the relation $P_{\rm ER}^s + P_{\rm EW}^s =1$ holds for the optimal trade-off between speed and error. Only close to the minimum error, $\eta_E$, does it drop very sharply to zero, see Fig.~\ref{fig:speed-error} in appendix~\ref{app:appendix_E}. In a non-equilibrium steady state, flux magnitudes, and consequently also the speed, are determined by both maxima and minima in the free energy landscape, corresponding respectively to energy barriers and discrete states \cite{Yu2022}. Conversely, the error rate, cost and entropy production rate per product are set by ratios of fluxes, which only depend on the energy barriers (kinetic control). As a result, the speed can be decoupled from the trade-off and varied independently by tuning the energy minima, which is, in fact, what the Pareto optimisation achieves. Such decoupling has, e.g., been observed in \emph{Escherichia coli} isoleucyl-tRNA synthetase~\cite{Yu2020}. Henceforth, we will not consider marginalised trade-offs involving the speed directly, but instead focus on the error-entropy production and error-cost trade-offs in the different regimes, which are shown in Fig.~\ref{fig:Paretogrid}. We discuss the different regimes and associated trade-offs in the following subsections.

\subsection{ERPR regime}
\begin{figure}[htp]
    \centering
    \includegraphics[width=\linewidth]{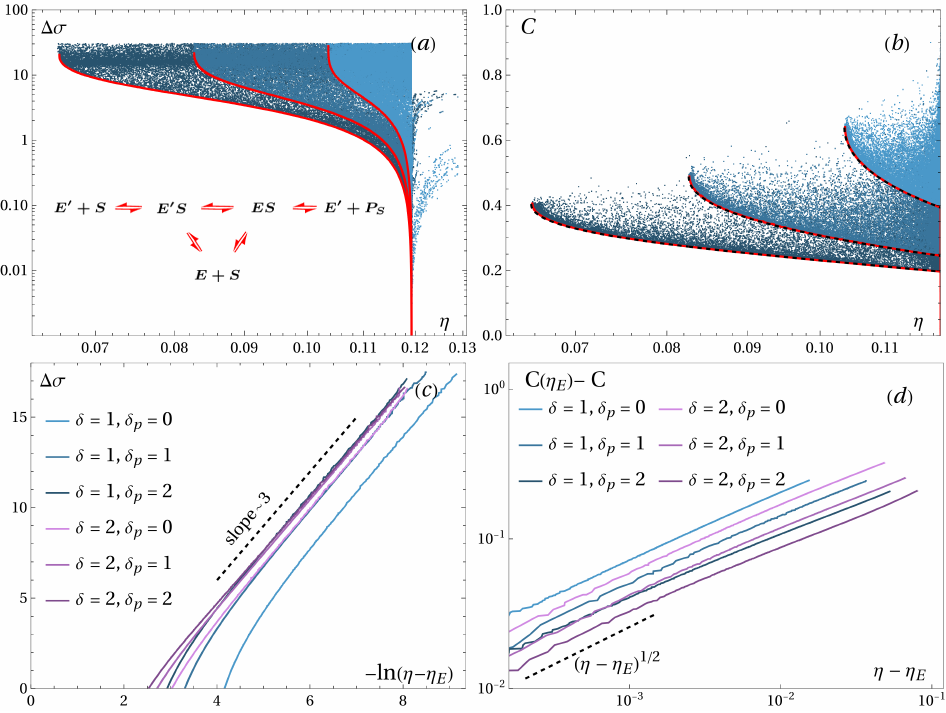}
    \caption{{\bf (a, b)} Projected Pareto fronts onto the $(\eta,\Delta\sigma)$ and $(\eta,C)$ subspaces for the $\gamma =2$, $\delta = 1$ ERPR regime shown in Fig.~\ref{fig:Paretogrid}, together with the marginalised trade-offs (red lines). Inset in (a): the preferred flow direction (red arrows) of the ERPR regime. In (b), the dashed black lines show the exact expression~\eqref{eq:cost_error_tradeoff}, which coincide perfectly with the numerically marginalised fronts. {\bf (c)} Entropy production according to the scaling law~\eqref{eq:sigma_eta_scaling}. {\bf (d)} Cost as a function of $\eta-\eta_E$, showing that for $\eta \gtrsim \eta_E$ the cost scales as $C\sim-(\eta-\eta_E)^\frac{1}{2}$.}
    \label{fig:ERPR_scaling}
\end{figure}

Considering now only the marginalised Pareto fronts onto the $(\eta,\Delta\sigma)$ and $(\eta,C)$ subspaces, it becomes clear that the discrimination regime influences the details of those trade-offs, as shown in Fig.~\ref{fig:Paretogrid}(b-c) for the ERPR regime where $\delta \leq \gamma$. The net preferred flow in the ERPR regime is shown in the inset in Fig.~\ref{fig:ERPR_scaling}(a).

For this regime, the proofreading cost $C$ jumps discontinuously from zero to a finite positive value at $\eta_{\rm eq}$. It signifies an abrupt shift in the optimal rate configuration when transitioning from the $\eta \gtrsim \eta_{\rm eq}$ regime to the $\eta \lesssim \eta_{\rm eq}$ regime. Similar behaviour is seen in a model of MM discrimination with dissipative resetting~\cite{Yu2022}, where a resetting cycle is driven by the cycle chemical potential difference, at the cost of increased entropy production. When $\delta$ increases to $\delta=\gamma$, the transition in proofreading cost becomes continuous at $\eta=\eta_{\rm eq}$, see Fig.~\ref{fig:Paretogrid}(b). 

The entropy production rate per product formed increases sharply from zero at $\eta = \eta_{\rm eq}$, where effective discrimination requires no free energy dissipation but relies solely on the difference in Gibbs free energy between the products, to infinity for small errors at $\eta\downarrow\eta_E$, where an infinite amount of free energy dissipation is required to maintain the minimal error rate. At a fixed error rate, increasing the proofreading parameter $\delta_p$ progressively lowers the minimal required proofreading cost. Concomitantly, the minimal entropy production rate is also lowered; a system with a better proofreading discrimination mechanisms requires less free energy to operate.

Based on the expression derived for MM discrimination (see appendix~\ref{app:appendix_D}), we find that close to the minimal error rate $\eta_E$, the entropy production-error trade-off follows a logarithmic scaling law,
\begin{equation}
    \label{eq:sigma_eta_scaling}
    \Delta\sigma\sim \alpha\,\ln{\left(\frac{1}{\eta-\eta_E}\right)}\,.
\end{equation}

Conversely, the cost-error trade-off is given by the exact relation~\eqref{eq:cost_error_tradeoff}, which asymptotically for $\eta\downarrow\eta_E$ exhibits a square root scaling, i.e.,
\begin{equation}
    \label{eq:cost_scaling}
    C - C(\eta_E)\sim  - (\eta - \eta_E)^\frac{1}{2}
\end{equation}

In Fig.~\ref{fig:ERPR_scaling}, we show the marginalised entropy production-error and cost-error trade-offs (red lines in (a,b)), along with the rescaled marginalised Pareto fronts for some parameter combinations according to equations~\eqref{eq:sigma_eta_scaling} and~\eqref{eq:cost_scaling}. Good data collapse is observed for both trade-offs according to the above scaling laws. Linear fits show that in the ERPR regime $\alpha = 3.01\pm0.01$ close to $\eta_{E}$, and that the square root scaling of equation~\eqref{eq:cost_scaling} can be retrieved numerically. 

Furthermore, in Fig.~\ref{fig:ERPR_scaling}(b), we show that the marginalised cost-error trade-off obtained numerically from the Pareto front (red lines) coincides exactly with the analytically computed cost function~\eqref{eq:cost_error_tradeoff} (dashed black line, see appendix~\ref{app:appendix_C}), where the cost at $\eta_{\rm eq}$ and the cost at the minimal error $\eta_{E}$ are given by respectively equation~\eqref{eq:cost_forward_eq} and~\eqref{eq:cost_forward_minimal}.

From this analysis, it becomes clear that the scaling behaviour of the entropy production as a function of the error rate is universal for the ERPR regime. The only effect of the parameters $\gamma,\,\delta,\,\delta_p$ is to tune the cut-off of the scaling laws.

\subsection{ERPR + MM regime}
When $\gamma < \delta\leq2\gamma+\delta_p$, a new region emerges on the Pareto front: for $\eta\geq\eta_M$, the scheme does not use the proofreading side reactions for errors in the range $[\eta_M,\,\eta_{\rm eq}]$, since discrimination only occurs at the MM error rate. When the minimal MM error rate $\eta_M$ is reached, however, the proofreading cycle is used to push the error rate even lower. At this instance, the minimal cost required continuously increases from $C = 0$ at $\eta = \eta_{M}$ to $C = C(\eta_E)$ at $\eta = \eta_{E}$. A similar switching exists between different mechanisms for conformationally fluctuating enzymes with multiple parallel reaction pathways \cite{Kumar2016}. For error rates lower than $\eta_M$, the entropy production again decreases with increasing $\delta_p$. Conversely, for error rates higher than $\eta_M$, increasing $\delta_p$ in fact increases the entropy production, see Fig.~\ref{fig:Paretogrid}(d). 

\begin{figure}[tp]
    \centering
    \includegraphics[width=\linewidth]{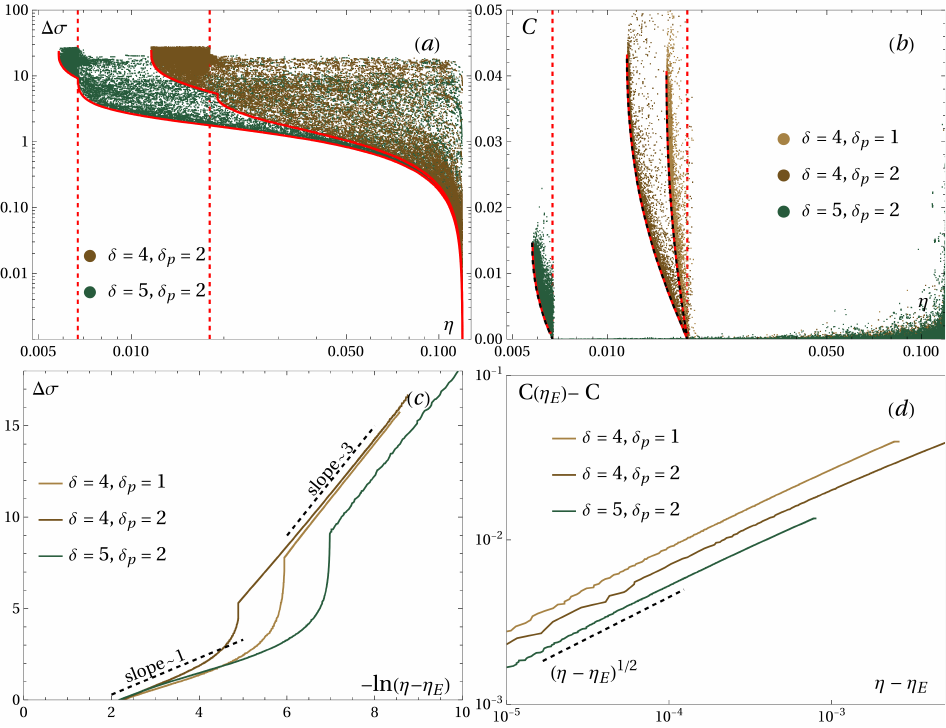}
    \caption{{\bf (a, b)} Projected Pareto fronts onto the $(\eta,\Delta\sigma)$ and $(\eta,C)$ subspaces for the $(\delta,\delta_p) = (4,2)$ and $(5,2)$ discrimination parameters in the ERPR + MM regime shown in Fig.~\ref{fig:Paretogrid}, together with the marginalised trade-offs (red lines). Red dashed vertical lines indicate $\eta_{\rm M}$. In (b), the dashed black lines show the exact expression~\eqref{eq:cost_error_tradeoff}, which coincide perfectly with the numerically marginalised fronts. {\bf (c)} Entropy production according to the scaling law~\eqref{eq:sigma_eta_scaling}. The cusps in the fronts indicate the transition from MM kinetics to ERPR. {\bf (d)} Cost as a function of $\eta-\eta_E$, showing that for $\eta \gtrsim \eta_E$ the cost scales as $C\sim-(\eta-\eta_E)^\frac{1}{2}$.}
    \label{fig:ERPRMM_scaling}
\end{figure}

The entropy production-error and cost-error trade-offs follow the same scaling laws as before, i.e., equation~\eqref{eq:sigma_eta_scaling} and~\eqref{eq:cost_scaling}, with $\alpha$ ranging from $\alpha \approx 1$ for $\eta \lesssim \eta_{\rm eq}$ to $\alpha \approx 3$ for $\eta\downarrow\eta_E$. Details are shown in Fig.~\ref{fig:ERPRMM_scaling}. The entropy production-error trade-off exhibits a phase transition at $\eta = \eta_{\rm M}$, where the optimal proofreading protocol changes \cite{Murugan2016,Seoane2016}. 
Similar phase transitions were observed for the Pareto optimal trade-off between the standard deviation and the mean of the dissipated work in a quantum dot \cite{Solon2018} and for the error-dissipation trade-off for a discriminatory network with correlations \cite{berx2023}.


\subsection{MM regime}
\begin{figure}[htp]
    \centering
    \includegraphics[width=\linewidth]{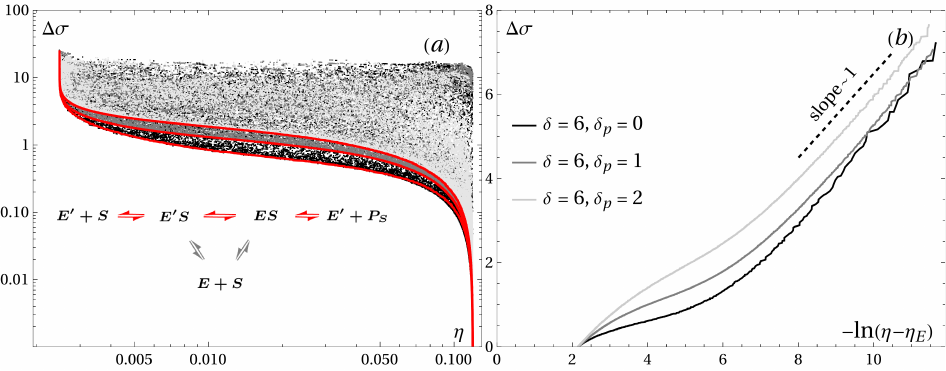}
    \caption{Relation between the entropy production per product $\Delta\sigma$ and the error rate $\eta$ in the MM regime, with $\delta = 6$ in (a). Red lines in (a) are numerically marginalised trade-off curves, which are replicated in (b) to show data scaling following equation \eqref{eq:sigma_eta_scaling}. Inset in (a): the preferred flow direction (red arrows) of the MM regime. }
    \label{fig:MM_scaling}
\end{figure}

Increasing $\delta$ or lowering $\delta_p$ to cross the boundary where $\delta = 2\gamma+\delta_p$ induces a transition to a fully MM discriminatory regime, where the proofreading kinetic parameter $\delta_p$ plays no role anymore in setting the minimal error rate. It is now fully determined by $\eta_M$ in the kinetic regime, which is a function of $\delta$ only. The proofreading side reaction is now never used in a useful fashion, since it cannot decrease the error any further. Increasing $\delta_p$ now has the adverse effect of raising the entropy production for a fixed error rate. In this regime, the entropy production per product increases smoothly from $\Delta\sigma = 0$ at $\eta=\eta_{\rm eq}$ to infinity at $\eta = \eta_{M}$, as is shown in Fig. \ref{fig:Paretogrid}(d-e-f). The average flow direction in the MM regime is shown in the inset in Fig. \ref{fig:MM_scaling}(a), where red arrows indicate pairs of transitions that occur more frequently.

The entropy production-error trade-off once again follows the same scaling laws as before, i.e., equation \eqref{eq:sigma_eta_scaling}, with $\alpha$ decreasing from $\alpha \approx 1$ close to $\eta \approx \eta_{\rm eq}$. For $\eta \downarrow \eta_{M}$, $\alpha$ increases again to $\alpha\approx 1$. The proofreading side chain is not used and the minimal cost is zero everywhere, so we do not show the cost-error trade-off.

\section{ERPR versus KPR \label{sec:ERPRvsKPR}}
Considering the mechanisms of the KPR and the ERPR, one can wonder about the realm of applicability of both discriminatory schemes. Comparing eq.~\eqref{eq:error_minimal} and eq.~\eqref{eq:error_minimal_KPR} for ERPR and KPR, respectively, it is clear that KPR always achieves the lowest minimal error rate for equal $\delta$, $\delta_p$ and $\gamma$. However, in order to make a fair comparison between the two schemes, we implement the KPR network as illustrated in Fig.~\ref{fig:networks}(b) with the associated free energy landscape shown in Fig.~\ref{fig:energy_landscape}. The proofreading rates $K_S^\pm$ in KPR are taken to be similar to the rates $\ell_S^\pm$ in the ERPR network, i.e., 
\begin{alignat}{2}
K_R^+ &= \omega_\ell \me^{\epsilon_p-\delta_p} &\qquad K_R^- &= \omega_\ell \me^{-\delta_p} \\
K_W^+ &= \omega_\ell \me^{\epsilon_p+\gamma} &\qquad K_W^- &= \omega_\ell\,.\nonumber
\end{alignat}
We also remove the constraint that $\epsilon_p \leq \epsilon_i$. Since the main difference between the two schemes is the proofreading transition, we fix all other rates to be the same as in the ERPR, as given by~\eqref{eq:rates}. As a result of the different modes of proofreading of both schemes, the proofreading parameter $\delta_p$ cannot be directly used in a similar fashion in both schemes, so we keep it as a free parameter that can be optimised in the Pareto algorithm. Conversely, since $\gamma$ and $\delta$ retain their physical interpretation of binding energy difference and initial kinetic binding barrier, respectively, they constitute the tuneable parameters that are common for both discriminatory schemes, and can be used for direct comparison. We again fix $\gamma=2$, $\epsilon=5$ and $\omega = 1$, such that $\delta$ is the only remaining control parameter. Additionally, we require that the value of the proofreading parameter is bounded by $\delta_p \leq \delta_{p,\rm max} = 5$, in order to maintain numerical accuracy and to avoid the minimal KPR error from becoming too small, since it can be seen from eq.~\eqref{eq:error_minimal_KPR} that $\eta_{K}\downarrow 0$ if $\delta_p$ can be increased indefinitely.

By varying $\delta$, the entropy production rate for a fixed error rate $\eta$ can be tuned; increasing $\delta$ leads to decreasing $\Delta\sigma$ for both the KPR and ERPR. The rate at which this decrease happens, however, is different for both schemes. For low values of $\delta$, there exists a critical error rate $\eta_c$ above which the ERPR scheme produces a smaller amount of entropy than the KPR. This effectively makes the energy relay a more effective proofreading scheme when the error rate is in the range $\eta\in[\eta_c,\eta_{\rm eq}]$. 

\begin{figure}[htp]
    \centering
    \includegraphics[width=0.9\linewidth]{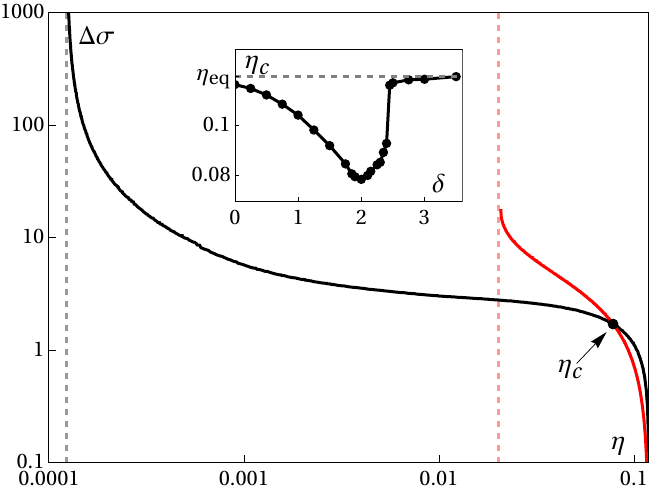}
    \caption{Comparison between the ERPR (red) and KPR (black) proofreading schemes, with $\gamma = \delta = 2$ and $\delta_p \in [0,5]$. Vertical lines correspond to the respective minimal error rates of both schemes. {\bf Inset:} Critical error rate $\eta_c$ above which $\Delta\sigma_{\rm KPR} \geq \Delta\sigma_{\rm ERPR}$, as a function of $\delta$. The error $\eta_{\rm eq}$ is indicated by a dashed line.}
    \label{fig:ERPR_vs_KPR}
\end{figure}

In Fig.~\ref{fig:ERPR_vs_KPR}, the error-entropy production trade-offs for both proofreading schemes are shown together. While the minimal errors differ in over two order of magnitude, the ERPR outperforms the KPR in the low-dissipation regime. The inset shows the critical error rate $\eta_c$ where KPR becomes energetically favourable to ERPR for a given value of $\delta$. When $\delta$ is zero, $\eta_c$ lies very close to $\eta_{\rm eq}$, and therefore KPR essentially always outperforms ERPR. However when $\delta$ increases towards $\delta = \gamma$ the error range wherein the ERPR is the preferable proofreading mechanism increases. This effect quickly vanishes when the system is not in the full ERPR regime $(\delta\leq\gamma)$ anymore; when $\delta \gtrsim \gamma$, the critical error $\eta_c$ sharply increases to $\eta_{\rm eq}$, such that the KPR scheme is the most effective for all error rates. This behaviour is robust against changing the range in which $\delta_p$ is allowed to vary. No noticeable changes to the inset of Fig.~\ref{fig:ERPR_vs_KPR} were observed when changing the maximum allowed value of $\delta_p$ in the range $\delta_{p,\rm max} \in [3,7]$.

\section{Conclusions \label{sec:conclusions}}
We have done an extensive thermodynamic analysis of Hopfield's energy relay proofreading scheme. This analysis reveals several crucial new insights that might have profound biological implications. Firstly we found that the energy-relay scheme consists of three distinct operational regimes: i) an energy relay regime, wherein discrimination occurs solely on the basis of the energy-relay mechanism, ii) a mixed relay-MM regime, where the discrimination can be performed by means of MM kinetics, until the minimal error is reached, after which the relay is activated and iii) a Michaelis-Menten regime, where the relay is not used and the mechanism reduces to known MM kinetics. Remarkably, each of these regimes has the same logarithmic divergence in the entropy production when the error rate approaches its minimal value. 

Surprisingly, we also found that the mixed regime features a phase transition, where the optimal kinetic rates change drastically. It would therefore be interesting to study the biological and evolutionary consequences of this type of phase-transitions in the Pareto front. Furthermore, it would be interesting to see whether other biological systems have similar types of phase-transitions.

Finally, we addressed the question of whether ERPR can thermodynamically outperform KPR in any regime. The answer to this turns out to be `yes'. Although it is generally possible to reach lower error rates with KPR in the high-dissipation limit, ERPR reaches lower error-rates when the amount of dissipation is limited. It would be interesting to see whether, e.g., the discrimination mechanism found for the CCA-adding enzyme shows this type of behaviour.



\begin{acknowledgments}
We thank S. Mondal and S. Pigolotti for their detailed comments on the use of the optimisation algorithm.

J.B. is funded by the European Union’s Horizon Europe framework under the Marie Sk\l odowska-Curie grant agreement No. 101104602 `STBR'. K.P. is funded by the European Union’s Horizon 2020 research and innovation program under the Marie Sk\l odowska-Curie grant agreement No. 101064626 `TSBC’, and from the Novo Nordisk Foundation (grant No. NNF18SA0035142 and NNF21OC0071284).
\end{acknowledgments}

\appendix

\section{Comparison of the minimal error rates}\label{app:appendix_B}
\subsection{Energy relay -- Michaelis-Menten}
To see that the ERPR scheme leads to a lower error than the MM scheme, we need to prove that $\eta_{E}\leq\eta_{M}$ for both the energetic and kinetic discrimination regimes. Let us start with the energetic regime, i.e., $\delta\leq\gamma$, where the MM error rate is equal to $\eta_{M} = 1/(1+\me^{\gamma})$. We need to prove that the following inequality holds:
\begin{equation}
   \frac{1}{1+\me^\gamma} \geq \frac{\me^{\gamma+\delta_p} + 2\me^{\gamma + \frac{\delta+\delta_p}{2}}-1-\me^\gamma - \me^\delta}{(1+\me^\gamma) (1+\me^\delta)(\me^{\gamma+\delta_p}-1)} = \eta_{E}\,.
\end{equation}
Since the denominator on the right-hand side of the inequality is strictly positive, we can cancel some terms, and rearrange the remaining ones as follows
\begin{equation}
    \me^{\gamma+\delta+\delta_p}\geq 2\me^{\gamma+\frac{\delta+\delta_p}{2}}-\me^\gamma\,.
\end{equation}
After rearranging the terms on one side and completing the square, the inequality reduces to
\begin{equation}
    \me^\gamma \left(1-\me^{\frac{\delta+\delta_p}{2}}\right)^2\geq0\,,
\end{equation}
which is true for all values of $\gamma,\,\delta,\,\delta_p$. In the kinetic discrimination regime where $\delta\geq\gamma$, the MM error rate is $\eta_{M} = 1/(1+\me^{\delta})$ and the inequality between the error rates becomes
\begin{equation}
    \eta_{M} = \frac{1}{1+\me^\delta} \geq \frac{\me^{\gamma+\delta_p} + 2\me^{\gamma + \frac{\delta+\delta_p}{2}}-1-\me^\gamma - \me^\delta}{(1+\me^\gamma) (1+\me^\delta)(\me^{\gamma+\delta_p}-1)} = \eta_{E}\,.
\end{equation}
Following the same procedure of rearranging the terms and completing the square, the inequality can be written as follows:
\begin{equation}
    \left(\me^\frac{\delta}{2}-\me^{\gamma+\frac{\delta_p}{2}}\right)^2\geq0\,,
\end{equation}
which is once again always true, proving our claim.

\subsection{Energy relay -- Kinetic proofreading}
Showing that $\eta_{K} \leq \eta_{E}$ is slightly trickier than the previous comparison. We again proceed with showing the inequality first in the energetic regime. Here, we need to show that the following holds:
\begin{equation}
    \frac{1}{1+\me^{2\gamma+\delta_p}} \leq \frac{\me^{\gamma+\delta_p} + 2\me^{\gamma + \frac{\delta+\delta_p}{2}}-1-\me^\gamma - \me^\delta}{(1+\me^\gamma) (1+\me^\delta)(\me^{\gamma+\delta_p}-1)}\,.
\end{equation}
Collecting and factoring both expressions to the l.h.s. of the equation, we simplify to
\begin{equation}
    \me^\frac{\delta}{2} +2 \me^\frac{\delta_p}{2} - \me^{\frac{\delta}{2}+\delta_p} - 2\me^{\gamma+\delta_p} \left(\me^\frac{\delta}{2}+\sinh{\left(\frac{\delta_p}{2}\right)}\right) \leq 0\,,
\end{equation}
which can be rewritten in the following compact form:
\begin{equation}
    \me^\gamma \geq \frac{2 \me^\frac{\delta_p}{2}+\me^\frac{\delta}{2}-\me^{\frac{\delta}{2}+\delta_p}}{2 \me^{\delta_p} \left(\me^\frac{\delta}{2}+\sinh{\left(\frac{\delta_p}{2}\right)}\right)}\,.
\end{equation}
It can now be seen that the l.h.s. is only a function of $\gamma$, which is bounded from below by unity. The r.h.s., however, depends on both $\delta$ and $\delta_p$. It can be checked that the r.h.s. exhibits a boundary global maximum at $\delta = \delta_p = 0$, which is equal to unity. Hence, since the l.h.s. is bounded from below by unity and the r.h.s. from above, $\eta_K \leq \eta_E$ is proven. 

For the kinetic regime, we follow the same procedure. Starting from
\begin{equation}
    \frac{1}{1+\me^{\gamma+\delta+\delta_p}} \leq \frac{\me^{\gamma+\delta_p} + 2\me^{\gamma + \frac{\delta+\delta_p}{2}}-1-\me^\gamma - \me^\delta}{(1+\me^\gamma) (1+\me^\delta)(\me^{\gamma+\delta_p}-1)}\,,
\end{equation}
we simplify to the following compact form
\begin{equation}
    \me^\gamma \geq \frac{\me^\delta + \me^\frac{\delta+\delta_p}{2}(2+\me^\delta)}{1+2 \me^\delta +\me^\frac{\delta+\delta_p}{2}}\me^{-\delta_p}\,,
\end{equation}
where, once again, the l.h.s. only contains $\gamma$. The r.h.s., however, is now increasing as a function of $\delta$ and decreasing as a function of $\delta_p$. Since at $\delta = 2\gamma + \delta_p$, $\eta_E = \eta_M$, the relevant global boundary maximum of the r.h.s. is located at $(\delta,\delta_p) = (2\gamma,0)$, such that the condition on $\gamma$ reduces to
\begin{equation}
    \me^\gamma \geq \frac{2+\me^\gamma + \me^{2\gamma}}{1+\me^\gamma + 2\me^{2\gamma}}\me^\gamma\,,
\end{equation}
which simply yields the trivial condition $\me^{2 \gamma} \geq 1$, proving our claim.

\section{Cost-error trade-off}\label{app:appendix_C}
The Pareto optimal proofreading cost $C$ as a function of the error rate $\eta$ is given by solving for $x$ in equation \eqref{eq:error_forward} and subsequently plugging the result into equation \eqref{eq:cost_forward}. This then becomes the following:
\begin{widetext}
    \begin{equation}
    \label{eq:cost_error_tradeoff}
    \begin{split}
        C(\eta) &= \frac{4 \me^{\gamma+\delta_p} f^{-2}(\gamma)\left(1-\eta f^{-1}(\delta)\right)^2}{\left(z+f^{-1}(\delta) -2f^{-1}(2\gamma+\delta_p)-\sqrt{(\eta-\eta_+)(\eta-\eta_-)}\right)\left(z-f^{-1}(\delta)+\sqrt{(\eta-\eta_+)(\eta-\eta_-)}\right)}\times\\
        &\left(1-\frac{f(\delta) f(\gamma)}{2 f(\delta-\gamma-\delta_p) (1-\eta f^{-1}(\delta))}\left(z+f^{-1}(\delta)+2(f^{-1}(\gamma+\delta_p)-1)\right)\right)\,,
    \end{split}
\end{equation}
\end{widetext}
with 
\begin{equation}
    \label{eq:error_roots}
    \eta_\pm = \frac{f^{-1}(\gamma+\delta_p)-f^{-1}(\delta)-f^{-1}(\gamma)\pm 2\me^{\gamma + \frac{\delta+\delta_p}{2}}}{\me^{\gamma+\delta_p}-1} f(\gamma) f(\delta)\,,
\end{equation}
and 
\begin{equation}
    \label{eq:z}
    z(\eta) = \left(1-\me^{\gamma+\delta_p}\right) + f^{-1} \left(1-\eta f^{-1}(\delta) \left(1-\me^{\gamma+\delta_p}\right)\right)\,.
\end{equation}

\section{Scaling of the entropy production-error trade-off}\label{app:appendix_D}
For a simple MM scheme with only one intermediate bound state, as in Fig.~\ref{fig:energy_landscape}(a), the trade-off between the error and entropy production rate can be found analytically in closed form. The transition matrix ${\bf \Gamma}$ is given by
\begin{equation}
    \label{eq:matrix_MM}
    {\bf \Gamma} = \begin{pmatrix}
        -(k_R^- + F) & k_R^+ & 0 \\
        k_R^- + F& -(k_R^+ + k_W^+) & k_W^- + F \\
        0 & k_W^+ & -(k_W^- + F)\,,
    \end{pmatrix}
\end{equation}
with kinetic rates given by equation~\eqref{eq:rates}. The error rate and entropy production rate per product are given by
\begin{equation}
    \label{eq:MM_error_rate_full}
    \eta = \frac{(k_R^- + F) k_W^+}{(k_R^- + F) k_W^+ + (k_W^- + F) k_R^+}\,.
\end{equation}
Solving for $F$ and inserting in the expression for the entropy production per product, given by equation~\eqref{eq:entropy_production}, the relation between $\eta$ and $\Delta\sigma$ is given by
\begin{equation}
    \label{eq:MM_entropy_full}
    \begin{split}
    \Delta\sigma &= \eta \ln{\left(\frac{k_W^+}{k_W^-}\frac{(k_R^- - k_W^-)  (1-\eta)}{\eta (k_R^+ + k_W^+) -k_W^+}\right)} \\
    &+ (1-\eta) \ln{\left(\frac{k_R^+}{k_R^-}\frac{(k_R^- - k_W^-)  \eta}{\eta (k_R^+ + k_W^+) -k_W^+}\right)}\,.
    \end{split}
\end{equation}
Inserting the expression for the kinetic rates based on the free energy landscape and rewriting results in the following
\begin{equation}
    \Delta\sigma = \eta \ln{\left(\frac{1-\eta}{1-\eta_{\rm eq}}\frac{\eta_{\rm eq}}{\eta}\right)} + \ln{\left(\frac{\eta_{\rm eq}-\eta_M}{\eta-\eta_M}\frac{\eta}{\eta_{\rm eq}}\right)}\,,
\end{equation}
such that for $\eta \downarrow \eta_M$, the entropy production rate scales logarithmically as
\begin{equation}
    \Delta\sigma\sim \ln{\left(\frac{1}{\eta-\eta_M}\right)}\,.
\end{equation}
As a result, when the amplitude of the scaling equals unity, the discriminatory scheme functions as a MM discriminatory network. Deviations in the amplitude signify non-MM behaviour. 

\section{speed-error trade-off}\label{app:appendix_E}
\begin{figure}[htp]
    \centering
    \includegraphics[width=0.9\linewidth]{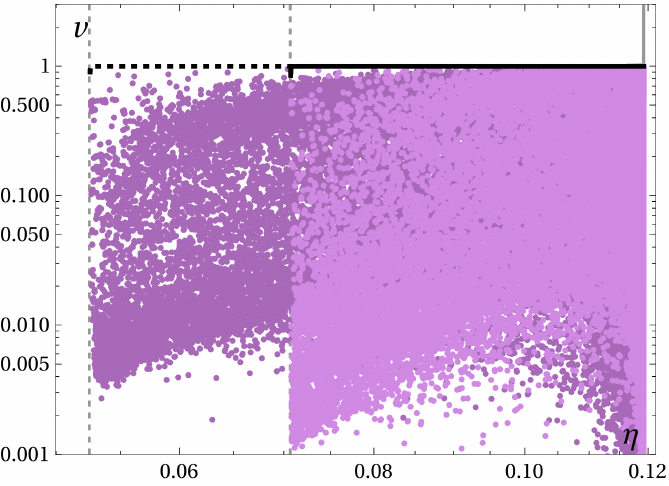}
    \caption{Speed-error trade-off for $\delta = 2$ and $\delta_p = 0$ (pink), $\delta_p = 1$ (purple). The black lines show the marginalised trade-off, indicating that the speed decouples from the error. Dashed vertical lines indicate the minimal error $\eta_E$.}
    \label{fig:speed-error}
\end{figure}

Fig.~\ref{fig:speed-error} shows the speed-error trade-off for $\delta = 2$. It can be seen that the speed completely decouples from the error, for all values of $\eta$, and steeply drops to zero at $\eta = \eta_E$.

\bibliographystyle{apsrev4-2}
\bibliography{biblio.bib}
\end{document}